\documentclass[aps,floatfix,prd,nofootinbib,superscriptaddress,reprint,showpacs,10pt,preprintnumbers,longbibliography]{revtex4-1}
\usepackage[utf8]{inputenc}
\usepackage{graphicx}
\usepackage{float}
\usepackage{amsmath}
\usepackage{amssymb}
\usepackage{mathtools}
\usepackage{siunitx}
\usepackage{amsfonts}
\usepackage{dsfont}
\usepackage{array}
\usepackage{bm}
\usepackage{mathrsfs}
\usepackage{pifont}
\usepackage{multirow}
\usepackage{upgreek}
\usepackage[dvipsnames]{xcolor}
\usepackage[pdftex,
  pdftitle={Symanzik Effective Hamiltonian for Lattice Quantum Field Theories},
  pdfauthor={Fernanda Steffens and Carsten Urbach},
  bookmarks,
  colorlinks,
  linkcolor=myblue,
  citecolor=mymagenta,
  menucolor=black,
  urlcolor=myblue,
  plainpages=false,
  pdfpagelabels,
  hypertexnames=false]{hyperref}
\usepackage{verbatim}
\usepackage{slashed}
\usepackage{cleveref}
\usepackage{ucs}
\usepackage{subfigure}
\usepackage{longtable,booktabs}
\usepackage{braket}

\newcommand{\tr}[1]{\mathrm{tr}\left(#1\right)}

\definecolor{mymagenta}{RGB}{200, 0, 100}
\definecolor{myblue}{RGB}{45, 48, 146}

\graphicspath{{plots/}}

\begin{document}
\title{Symanzik Effective Hamiltonian for Lattice Quantum Field Theories}
\author{Fernanda Steffens}
\affiliation{Helmholtz-Institut für Strahlen- und Kernphysik, University of Bonn, Nussallee 14-16, 53115 Bonn, Germany}
\affiliation{Bethe Center for Theoretical Physics, University of Bonn, Nussallee 12, 53115 Bonn, Germany}
\author{Carsten Urbach}
\affiliation{Helmholtz-Institut für Strahlen- und Kernphysik, University of Bonn, Nussallee 14-16, 53115 Bonn, Germany}
\affiliation{Bethe Center for Theoretical Physics, University of Bonn, Nussallee 12, 53115 Bonn, Germany}

\date{\today}
\begin{abstract}
  We propose a framework to construct the Symanzik effective
  Hamiltonian for lattice quantum field theories directly based on
  dimensional analysis and symmetries in the Hamiltonian
  formulation. This allows to classify all operators which can
  contribute at a given order in the spatial lattice spacing. This
  list can be reduced using the equations of motion and other
  identities. We exemplify the framework for $\phi^4$ and pure gauge
  theory, obtaining results consistent with earlier publications for
  the case of classical improvement. It
  provides the means to develop a non-perturbative improvement
  programme in the Hamiltonian formulation of lattice quantum field
  theories in the future.
\end{abstract}

\maketitle

\section{Introduction}

Since its invention by Wilson in 1974, lattice quantum chromodynamics
(LQCD) has become the main non-perturbative method to investigate QCD
from first principles. Monte Carlo simulations are performed based on
the Euclidean action regularised on a finite space-time
lattice. Besides algorithmic progress, the non-perturbative
Symanzik improvement programme based on seminal work by
Symanzik~\cite{Symanzik:1981hc,Symanzik:1983gh,Symanzik:1983dc}, 
Lüscher and Weisz~\cite{Weisz:1982zw,Luscher:1984xn,Luscher:1984is,Luscher:1996sc} was
instrumental for a precision of below one 
percent possible nowadays for some observables.

The Hamiltonian formalism for LQCD, or more generally lattice gauge
theories, has been developed in parallel since the paper by Kogut and
Susskind~\cite{Kogut:1974ag}. 
More recently, developments in tensor-network methods and quantum
 computing have renewed interest in the Hamiltonian formulation.
Classical improvement of lattice gauge Hamiltonians has been studied 
both through constructions based on improved lattice actions~\cite{Luo:1998dx}  
and directly within the Hamiltonian
formulation~\cite{Carlsson:2001wp}, see also
Refs.~\cite{Carena:2022kpg,Illa:2025njz,Illa:2025dou}.  
A related work~\cite{Grosse-Knetter:1993tae} by Grosse-Knetter
addresses the perturbative elimination of higher order time
derivatives in effective Lagrangians and their Hamiltonian
formulation. 

In this paper we propose a framework for constructing a Symanzik
effective Hamiltonian directly based on the symmetries of the lattice
theory. 
It provides the Hamiltonian counterpart of Symanzik’s effective action. 
Contributions to the effective Hamiltonian are classified order by order in the 
spatial lattice spacing $a_s$ using dimensional analysis and the symmetries of the theory. 
Independent on-shell operator bases are obtained by identifying redundancies, 
including those generated by commutators with the leading continuum Hamiltonian. 
We also discuss the connection to Symanzik’s effective action through 
the temporal continuum limit and a perturbative Legendre transformation.

We apply the framework to $\phi^4$ theory and pure lattice gauge theory, 
constructing the operator bases through order $a_s^2$. 
For the scalar theory, we establish an explicit correspondence with the effective-action 
coefficients and compare with Symanzik’s results, and check the construction for 
the tree-level expansion of the Hamiltonian.
For the gauge theory, we discuss the reduction of the effective-action 
operators to the Hamiltonian basis, and compare against the known 
quadratic tree-level corrections. 

This work is organised as follows. In \cref{sec:Hamiltonian} we
introduce the Symanzik effective Hamiltonian. In \cref{sec:Legendre}
we argue how this effective Hamiltonian can be derived from the
Symanzik effective action on anisotropic lattices via a Legendre
transformation. While we use $\phi^4$ theory as the example in those
two sections, \cref{sec:Gauge} is devoted to the derivation of the
effective Hamiltonian at order $a_s^2$ for pure lattice gauge
theories.

\section{Symanzik Hamiltonian}
\label{sec:Hamiltonian}

We will start by defining the effective Hamiltonian equivalent to the
Symanzik effective action. We will later give an argument for why such
an effective Hamiltonian should exist. The effective Hamiltonian 
represents a low energy Hamiltonian valid for physics well below
the cut-off energy $\sim1/a_s$. In $d$ spatial dimensions, this effective
Hamiltonian reads 
\begin{equation}
  H_\mathrm{eff} = H_0 + a_s\, H_1 + a_s^2 H_2 + \ldots
\end{equation}
following a similar notation as was used in
Ref.~\cite{Luscher:1996sc}. Since we are in the Hamiltonian
formulation, only the spatial lattice spacing $a_s$ is relevant.
This Hamiltonian $H_\mathrm{eff}$ acts on
the continuum Hilbert or Fock space, so $H_0$ is the continuum
Hamiltonian regulated by a much smaller spatial lattice spacing
$\tilde a_s\ll a_s$
(Symanzik used dimensional regularisation instead)
obtained from the Hamiltonian density $\mathcal{H}_0(x)$ via
\begin{equation}
  H_0\ =\ \int\mathrm{d}^d x\ \mathcal{H}_0(x)\,.
\end{equation}
The correction terms are then defined as
\begin{equation}
  H_k\ =\ \int\mathrm{d}^d x\ \sum_i c_{k,i}\,\mathcal{O}_{k,i}(x)\,.
\end{equation}
The densities $\mathcal{O}_{k,i}$ are of appropriate dimension $d+1+k$,
and represent combinations (monomials) of canonical fields $\phi$ and
momenta $\pi$ with commutation relations reading
\begin{equation}
  \label{eq:fieldComm}
  [\phi(x), \pi(y)]\ =\ i\, \delta(x-y)\,.
\end{equation}
In particular, the $\mathcal{O}_{k,i}$ are only allowed to contain
spatial derivatives, no derivatives with respect to time, see also the
next section. Like in
Ref.~\cite{Luscher:1996sc}, powers of the mass are 
included in the densities $\mathcal{O}_{k,i}$ and, therefore, the
counting of the dimensions. The coefficients $c_{k,i}$ are
dimensionless functions of the coupling, beyond tree-level they also
carry a logarithmic $a_s$ dependence.

At fixed $k$, the list of possible $\mathcal{O}_{k,i}$ is restricted
by gauge invariance and the exact symmetries of the lattice
Hamiltonian, for instance spatial lattice symmetries and charge
conjugation. Exact symmetries of the lattice Hamiltonian must be
respected order by order in the lattice spacing $a_s$
\begin{equation}
  U\, H_\mathrm{eff}\, U^{-1}\ =\ H_\mathrm{eff}\,,
\end{equation}
with $U$ the corresponding symmetry operator. Time reversal relates
coefficients to their complex conjugates by acting via complex
conjugation. Since we are working in the Hamiltonian formalism, boosts
are not available as a symmetry (time is continuous, space is
discretised): terms typically related by Lorentz invariance obtain
independent coefficients as long as they are invariant under the point group
$O_h$. The breaking of Lorentz invariance is particularly important
because it introduces corrections to the speed of light which can be
relevant at order $(a_s)^0$. 

The list can be further reduced by noting that total spatial
derivatives cancel via the integral over space. Like in the case of
the Symanzik action also for the Hamiltonian equations of motion can
be used. For the Hamiltonian case time evolution in the continuum is
driven by 
\begin{equation}
  \dot \Phi = i\,[H_0, \Phi]
\end{equation}
for a local hermitian operator $\Phi$ with no explicit time
dependence. For instance for $\Phi$ of
dimension one, which generates the infinitesimal
transformation 
\begin{equation}
  e^{i a_s \Phi}\ H_\mathrm{eff}\ e^{-i a_s \Phi} = H_0 + a_s(H_1 +
  i[\Phi, H_0]) + \ldots\,,
\end{equation}
which redefines $H_1$ by the time derivative of $\Phi$. The general rule
would be to chose a $\Phi_k$ with dimension $[\Phi_k]=k$ to modify
$H_k$. Note that valid generators $\Phi$ must be gauge invariant under
local and global gauge transformations.

In addition to the Hamiltonian, also any lattice operator is a source
of lattice artefacts. Those may be written as
\begin{equation}
  O\ =\ Z_O\left(O_0 + \sum_k a_s^k\sum_i d_{k,i}\, O_{k,i}\right)\,,
\end{equation}
with an appropriately chosen renormalisation factor $Z_O$. The
$O_{k,i}$ have dimensions $[O_0] + k$ and transform under the lattice
symmetries exactly in the way $O$ does. Note that while the spectrum
of $H$ is conserved by the $\exp(ia_s\Phi)$ transformation from above,
the $d_{k,i}$ change under such unitary transformations.

Corrections can now be computed in perturbation theory, for instance
for energy levels $E_n$ in non-degenerate perturbation theory
\begin{equation}
  \begin{split}
    E_n = E_n^0& + a_s\bra{n^0}H_1\ket{n^0}\\
    &+
    a_s^2\left[\bra{n^0}H_2\ket{n^0}
     + \sum_{m\neq
       n}\frac{|\bra{m^0}H_1\ket{n^0}|^2}{E_n^0-E_m^0}\right]\\
    &+ \ldots\\
  \end{split}
\end{equation}
Likewise, the wave functions receive corrections systematically
\begin{equation}
  \begin{split}
    \ket{n} = &\ket{n^0}
    +a_s\sum_{m\neq n}\ket{m^0}\frac{\bra{m^0}H_1\ket{n^0}}{E_n^0-E_m^0}\\
    &+ O(a_s^2)\,.
  \end{split}
\end{equation}

\subsection{Scalar $\phi^4$ Theory}

We work on a $d=3$ spatial lattice with sites
\begin{equation}
  n\in\Lambda=\{0,\dots,N_s-1\}^3
\end{equation}
with periodic boundary conditions.
The lattice Hamiltonian reads
\begin{equation}
  H_{\rm lat}=\sum_{n\in\Lambda}
  a_s^3\Big[\tfrac12\pi_n^2+\tfrac12 c_0^2\sum_i\Big(\frac{\phi_{n+e_i}-\phi_n}{a_s}\Big)^2
    + V(\phi)\Big]\,. 
\label{eq:lattice_hamiltonian}
\end{equation}
with
\begin{equation}
  V(\phi_n) = +\tfrac12m_0^2\phi_n^2+\tfrac{\lambda_0}{4!}\phi_n^4
\end{equation}
Due to symmetries reduced to $O_h$, the term $\pi^2$ and the gradient
part in the lattice Hamiltonian renormalise independently. This fact
is reflected in the additional coefficient $c_0$, 
which is tuned so that the low-momentum dispersion relation yields 
the physical propagation speed. At tree level this corresponds to $c_0=1$

The continuum Hamiltonian reads
\begin{equation}
  H_0 = \int\mathrm{d}^3 x \left(\frac{1}{2}\pi^2 + \frac{1}{2}(\nabla
  \phi)^2 + \frac{1}{2}m^2\phi^2 + \frac{\lambda}{4!}\phi^4\right)\,.
\end{equation}
As symmetries we have available spatial translations, $O_h$,
$(\phi,\pi)\to(-\phi,-\pi)$ ($\mathbb{Z}_2$ symmetry), and
$(\phi,\pi)\to(\phi,-\pi)$ (time reversal $\mathcal{T}$). Dimensions
are
\begin{equation}
  [\phi]=\tfrac{d-1}{2}\,,\qquad [\pi]=\tfrac{d+1}{2}\,,\qquad [\partial_i]=1\,.
\end{equation}
Thus, for $H_1$ in the effective Hamiltonian possible combinations of
dimension $d+2=5$ are $\phi\partial\partial\partial\phi$,
$\pi\partial\partial\phi$, $\pi\partial\pi$, $\phi^3\pi$,
$\phi^2\partial\phi^2$, $\phi^5$, and $\pi^2\phi$. An odd number of spatial
indices is forbidden by $O_h$, removing the first, the third, and the
fifth from the list. An odd number of $\pi$ is forbidden by $\mathcal{T}$,
removing the second and the fourth. Finally, $\mathbb{Z}_2$ removes
the last two on the list due to their odd number of $\phi$ fields.
Thus, this scalar field theory has no $O(a)$ correction in the
effective Hamiltonian consistent with the original work by
Symanzik~\cite{Symanzik:1983dc}. Note that the dimension five mass
weighted operators
$m\pi^2\,,\ m(\nabla\phi)^2\,,\ m^3\phi^2\,,\ m\phi^4$ are odd in $m$
and, therefore, cannot appear because the Hamiltonian depends on the
mass only via $m_0^2$. 

However, at $O(a_s^2)$ there is a list of allowed operators in Hermitian ordering
\begin{equation}
  (\nabla^2\phi)^2\,,\ \sum_i(\partial_i^2\phi)^2\,,\ (\nabla\pi)^2\,,\ \phi^2(\nabla\phi)^2,
  \pi\phi^2\pi, \phi^6
\end{equation}
of appropriate dimension. Different orderings differ by
$\delta(0)\times$(lower-dimension terms). 
In addition we can have combinations with
the mass $m$
\begin{equation}
  m^2\pi^2\,,\ m^2(\nabla\phi)^2, m^4\phi^2, m^2\phi^4\,.
\end{equation}
In this theory we can see how the freedom of adding time
derivatives to $H_2$ allows us to remove redundancies. Take for
instance
\begin{equation}
  \Phi_K = \frac{1}{2}\int\mathrm{d}^3 x(\phi\, K\,\pi + \pi\, K\,\phi)\,,
\end{equation}
with $K=-\nabla^2$. Using the fundamental commutation relations
\cref{eq:fieldComm}, one sees that
\begin{equation}
  \begin{split}
    i[H_0,\phi(x)] &= \pi(x)\,, \\
    i[H_0, \pi(x)] &= -(-\nabla^2 + m^2)\phi - \frac{\lambda}{6}\phi^3\,. \\
  \end{split}
\end{equation}
This can be used to derive
\begin{equation}
  \begin{split}
    i[H_0,\Phi_K] &=\\
    \int\mathrm{d}^3 x\ &\Big[(\nabla\pi)^2-(\nabla^2\phi)^2-m^2(\nabla\phi)^2-\tfrac{\lambda}{2}\,\phi^2(\nabla\phi)^2\Big]\,.
  \end{split}
\end{equation}
Thus, by adding $i[H_0,\Phi_K]$ with an appropriate coefficient to
$H_2$ allows one to remove the redundant $(\nabla\pi)^2$ from the list
of operators. We remark that this transformation is equivalent to a
canonical transformation of the canonical fields $\phi,\pi$ with
\begin{equation}
  \phi\to\phi - c\,a_s^2\nabla^2\phi\,,\qquad \pi\to\pi + c\,a_s^2\nabla^2\pi\,,
\end{equation}
which preserves the commutation relations. The coefficient $c$ needs to
be fixed according to the coefficient of $(\nabla\pi)^2$.
Similarly, with
\begin{equation}
  \Phi = \frac{1}{2}\int\mathrm{d}^3 x\left(\phi^3\pi + \pi\phi^3 \right)
\end{equation}
one obtains up to a constant
\begin{equation}
  \begin{split}
    i[H_0,\Phi]&=\\
    \int\mathrm{d}^3 x &\Big[3\,\pi\phi^2\pi-3\,\phi^2(\nabla\phi)^2-m^2\phi^4-\tfrac{\lambda}{6}\phi^6\Big]\,,\\
  \end{split}
\end{equation}
which allows one to remove $\pi^2\phi^2$. One is left with the
following on-shell basis for $H_2$
\begin{equation}
  \big\{(\nabla^2\phi)^2,\ \textstyle\sum_i(\partial_i^2\phi)^2,\ \phi^2(\nabla\phi)^2,\ \phi^6\big\}
\label{eq:basis}
\end{equation}
plus (marginal) terms proportional to factors of the mass. Note that
with
\begin{equation}
  \Phi_2 = \frac{1}{2}\int\mathrm{d}^3 x(\phi\pi + \pi\phi)\,,
\end{equation}
which has the following commutator with $H_0$
\begin{equation}
  i[H_0, \Phi_2] = \int\mathrm{d}^3 x \left[\pi^2 -(\nabla\phi)^2-m^2\phi^2-\frac{\lambda}{6}\phi^4\right]\,,
\end{equation}
the $m^2\pi^2$ term can be removed as well with the unitary
transformation $\exp(i\epsilon a_s^2m^2\Phi_2)$, and $\epsilon$ chosen
accordingly. Therefore, to this order in $a_s$
none of the $H_2$-terms contains the canonical momentum $\pi$.

The above procedure is equivalent to using the equations of motion
$\ddot\phi+(-\nabla^2 + m^2)\phi+\tfrac\lambda6\phi^3=0$ in the Lagrangian
formalism to remove redundant operators. Notably, due to missing boost
symmetry $\pi^2$ and $(\nabla\phi)^2$ obtain independent
coefficients (the speed of light). At $O(a_s^2)$ and before reduction this implies
independent coefficients for the terms $m^2\pi^2$ and $m^2(\nabla\phi)^2$ as well.

\subsection{Tree-level matching}
\label{sec:tree}

Using the basis written in \cref{eq:basis}, we construct $H_2$
\begin{equation}
          \begin{aligned} 
     H_2=\int \mathrm{d}^3x\, \Bigl[ c_1(\nabla^2\phi)^2 + & c_2\sum_i(\partial_i^2\phi)^2 \\
             & +c_3\phi^2(\nabla\phi)^2 +c_4\phi^6 \Bigr]
           \end{aligned}
\label{eq:fourbasis}
\end{equation}
plus the remaining mass dependent terms.
The coefficients are determined via matching or renormalization conditions. 
We can fix the coefficients at tree level via an expansion of the discrete derivative in \cref{eq:lattice_hamiltonian}. 
There is no a priori requirement that all the $c_i$ are nonzero, but even  
if all of them would be zero, higher order loops can give them a finite value.

We match the lattice Hamiltonian \cref{eq:lattice_hamiltonian} to 
the on shell basis \cref{eq:basis}, keeping in mind that 
at tree level, $m_0=m$ and $\lambda_0=\lambda$. We then
expand $H_{\rm lat}$ in powers of $a_s$ for smooth fields and compare with $H_{\rm eff}$.
We work in momentum space, where the two quadratic operators of the
basis \cref{eq:basis} become
the independent polynomials $(\mathbf p^2)^2$ and $\sum_i p_i^4$, so that their coefficients
follow by comparing polynomials in $\mathbf p$. 

Let $H_\mathrm{grad}$ be the gradient term of \cref{eq:lattice_hamiltonian}. We Fourier transform the
fields in a periodic spatial box of length $L$ and then take
the infinite volume limit at fixed $a_s$, resulting in
\begin{equation}
 H_{\mathrm{grad}}=
 \frac{c_0^2}{2}\int_{\mathbf p}\phi(-\mathbf p)
 \hat{\mathbf p}^{\,2}\phi(\mathbf p),\qquad
  \label{eq:matchingkernel}
 \end{equation}
with
\begin{equation}
 \hat{\mathbf p}^{\,2}
 =\sum_{i=1}^{3} \frac4{a_s^2}\sin^2\!\left(\frac{a_s p_i}{2}\right)
 =\mathbf p^2-\frac{a_s^2}{12}\sum_{i=1}^{3} p_i^4
 +O(a_s^4p^6).
\end{equation}
The corresponding terms with two fields in $H_2$ are
\begin{equation}
 a_s^2\int_{\mathbf p}\phi(-\mathbf p)
 \left[c_1(\mathbf p^2)^2+c_2\sum_i p_i^4\right]
 \phi(\mathbf p).
 \label{eq:twokernel}
\end{equation}
The coefficients are determined by comparing the two expressions for $|a_s p_i|\ll1$.
Because $(\mathbf p^2)^2=\sum_i p_i^4+2\sum_{i<j}p_i^2p_j^2$ 
and the lattice kernel contains no $p_i^2p_j^2$ term at this order, $c_1=0$.
Matching the remaining $\sum_i p_i^4$ coefficient then gives
$c_2=-1/24$, for $c_0$ tuned to 1. 

The quartic term of \cref{eq:lattice_hamiltonian} is local and the Hamiltonian has no six-field
term, so neither produces an $O(a_s^2)$ contribution with four fields and two derivatives,
or with six fields. Thus $c_3$ and $c_4$ vanish at tree level.

Different equation of motion reductions or canonical field redefinitions
can redistribute coefficients across operator sectors, so those zeros also refer to the stated matching
convention.

In general, the  $c_i$ are also affected by a loop expansion.  
For example, to compute the loop contributions for the two field coefficients,
$c_1^{\rm loops}$ and $c_2^{\rm loops}$, we need the
dispersion relation following from the pole of the propagator,
\begin{equation}
  E(\mathbf p)^2-m_0^2 - c_0^2 \hat{\mathbf p}^2-\Sigma(E(\mathbf p)^2,\mathbf p;a_s,\lambda_0)=0,
\end{equation}
where $\Sigma$ is the self-energy, and the low momentum slope is
\begin{equation}
  c_{\rm pole}^2\equiv\left.\frac{\partial E^2}{\partial p^2}\right|_{p=0}.
\end{equation}
At tree level $\Sigma=0$ and $c_{\rm pole}^2=c_0^2$. 
At one loop the only contribution is a tadpole, with 
no external momentum flowing through the loop. As it is independent of $E$ and $\mathbf p$, the effect
is to shift the mass, leaving $c_{\rm pole}$ unchanged. At two loops the integral in the sunset diagrams 
depends on the external energy and momentum, and generates the first
dependence of $\Sigma$ on $E$ and $\mathbf p$. Consequently, the two-loop calculation can affect
both the coefficient $c_{pole}^2$ of the momentum square term and the $O(a_s^2p^4)$ term in the
dispersion relation, with the latter contributing to the matching of $c_1$ and $c_2$.
 
\section{The Lagrangian Route}
\label{sec:Legendre}

We return to the question of whether the effective Hamiltonian
should exist. 
The purpose is to relate the Lagrangian expansion to the 
Hamiltonian proposed in \cref{sec:Hamiltonian}. For an isotropic Euclidean
lattice, the Symanzik expansion has the form
\begin{equation}
  S = S_0 + a S_1 + a^2 S_2 +\ldots
\end{equation}
where $S_0$ represents the continuum action, or rather a regularised
version of it e.g. with a much smaller lattice spacing. The terms
$S_k$ with $k>0$ are corrections due to the lattice regularisation
assuming the isotropic lattice spacing $a=a_s=a_t$ to be sufficiently
small to resolve all relevant scales of the theory.

We will use here a slightly modified version already hinted at by
Symanzik in his first publication Ref.~\cite{Symanzik:1983dc}: the use
of an anisotropic formulation with temporal lattice spacing $a_t$ different
from the spatial one $a_s$\footnote{We thank Roberto Frezzotti for
pointing this out to us.}.
Assuming that the renormalized temporal continuum limit exists, we
take $a_t\rightarrow 0$ at fixed $a_s$. The remaining expansion is then
\begin{equation}
  S = S_0 + a_s S_1 + a_s^2 S_2 +\ldots
\label{eq:action_spatial_expansion}
\end{equation}
Each of the terms $S_k$ can be written as
\begin{equation}
  S_k = \int \mathrm{d}^4x\, \mathcal{L}_{E,k}(\phi,\dot\phi, \ddot\phi,
  \ldots)
\end{equation}
of Lagrange densities $\mathcal{L}_{E,0}, \mathcal{L}_{E,1},\ldots$ with mass
dimension $4+l$. We may write
\begin{equation}
  \label{eq:density}
  \mathcal{L} = \sum_{l=0}^\infty a_s^l\, \mathcal{L}_{l}\,,\quad   S = \int\mathrm{d}^4x\, \mathcal{L}_{E}\,.
\end{equation}
The contributions $\mathcal{L}_{E,k}$ at each order are then determined by
a dimensional analysis and the symmetries of the action, taking into
account that Lorentz invariance is broken due to the asymmetry between
time and space. We use Latin indices $i,j=1,\ldots,4$ for Euclidean
space components,  and Greek indices $\mu,\nu=0,\ldots,3$ for
Minkowski space-time components.  
Spatial sums are explicitly restricted to the three spatial directions. 
The continuation between signatures is defined by
\begin{equation}
  \tau =i t,\qquad
  \partial_\tau=-i\partial_0,\qquad
  \mathcal L_M
  =-\left.\mathcal L_E\right|{\partial\tau\to-i\partial_0}.
\end{equation}
The effective operators are initially classified in Euclidean space. Higher time derivatives in the 
correction terms are removed perturbatively before performing the ordinary Legendre transformation 
of the resulting Minkowski Lagrangian.

The next step involves the use of the equations of motion to remove
any higher order time derivative in the $\mathcal{L}_k, k\geq1$ up to
the order in $a_s$ one is interested in.

The transition to the Hamiltonian formalism in field theory is
performed by the step
\begin{equation}
H(\pi,\phi) = \int\mathrm{d}^3 x \left(\pi\dot\phi - \mathcal{L}_M(\phi,\dot\phi)\right)
\end{equation}
with
\begin{equation}
  \label{eq:pi}
  \pi = \frac{\partial \mathcal{L}_M}{\partial\dot\phi} = \frac{\partial
    \mathcal{L}_{M,0}}{\partial\dot\phi} + a_s \frac{\partial
    \mathcal{L}_{M,1}}{\partial\dot\phi} + \ldots\,.
\end{equation}
The Legendre transformation can now be performed
perturbatively. Writing $v=\dot\phi$ and the Hessian
$M=\partial^2\mathcal{L}_{M,0}/\partial v^2$ which we assume to be
invertible and $\pi=\partial\mathcal{L}_M /\partial v$, one obtains
\begin{equation}
  \mathcal{H} = \mathcal{H}_0 + \sum_ka_s^k\, \mathcal{H}_k\,.
\end{equation}
with
\begin{equation}
  \begin{split}
    \mathcal{H}_0 &= \pi v_0 - \mathcal{L}_0(\phi, v_0)\,,\\
    \mathcal{H}_1 &= -\mathcal{L}_{M,1}\,,\\
    \mathcal{H}_2 &= -\mathcal{L}_{M,2} +
    \frac{1}{2}(\partial_v\mathcal{L}_{M,1})M^{-1}(\partial_v\mathcal{L}_{M,1})\,,\\ 
  \end{split}
\end{equation}
up to corrections of order $a_s^3$, with all $\mathcal{L}_{M,k}, k=0,1,2$
evaluated at $v=v_0(\phi,\pi)$, the velocity of the unperturbed theory.
Canonical quantisation of $(\phi,\pi)$ then yields $H_\mathrm{eff}$,
up to ordering terms that renormalise lower-dimension coefficients.
This procedure can be continued to higher orders in $a_s$. Note that
since $\mathcal{H}$ is stationary in $v$, the correction to $v$ at
order $k$ never contributes to order $a_s^k$. 
The terms $\mathcal{H}_l$ have again mass dimension $4+l$.

Above we assumed that all higher order time derivatives can be removed
using the equations of motion and field
re-definitions, for the argument see the arguments of Lüscher and
Weisz in Ref.~\cite{Luscher:1984xn}, and also
Ref.~\cite{Georgi:1991ch}.

We remark that a way to deal
with higher order time derivatives in the Legendre transformation
is the so-called Ostrogradsky
construction~\cite{Ostrogradsky:1850fid}: for each additional order
in time derivatives a new pair of fields and canonical momenta is
introduced. It leads in general to the so-called Ostrogradsky
instability, see Ref.~\cite{Woodard:2015zca}: the Hamiltonian obtained
by this procedure is not necessarily bounded from below, or,
equivalently, states with negative norm appear in the spectrum. 
Such states are spurious, also called ghost states with masses of
order $1/a_s$ at the cut-off, see 
Ref.~\cite{Simon:1990ic}. These ghost states should not be treated as
dynamical~\cite{Jaen:1986iz,Eliezer:1989cr}.
See also Ref.~\cite{Jansen:1993jj} for a 
different, more lattice driven viewpoint. 

Another way to think of these states is in the context of reflection
positivity, see Refs.~\cite{Osterwalder:1973dx,Osterwalder:1974tc}. On
the lattice, second or higher order time derivatives couple lattice
sites of two or more in time direction. For the consequences for
reflection positivity in the Lagrangian formalism see
Ref.\cite{Luscher:1984is} (see also Ref.~\cite{Luscher:1976ms}).

With our proposed Symanzik effective Hamiltonian framework, these
problems are circumvented entirely by imposing that no time
derivatives of the canonical fields may appear in the effective
Hamiltonian.

\subsection{Scalar Field Theory revisited}

We first recall the isotropic Euclidean operator mass independent 
basis used by Symanzik 
in Ref.~\cite{Symanzik:1983dc}. In this case, $\mathcal{L}_{E,1}=0$, 
but for  $\mathcal{L}_{E,2}$ the following basis of operators contribute at
$O(a_s^2)$
\begin{equation}
\left\{\sum_{i=1}^{4}(\partial_i^2\varphi)^2,
(\Box_E\varphi)^2,
\varphi^2\sum_{i=1}^{4}(\partial_i\varphi)^2,
\varphi^6 \right\},
\label{eq:isotropic_basis}
\end{equation}
modulo total derivatives.
This basis is invariant under $\varphi\to-\varphi$ and the isotropic lattice hypercubic group. 
It contains two independent quadratic operators with four derivatives, one quartic operator 
with two derivatives, and one six-field operator. It has not yet been reduced using the equations of motion.
Explicit mass-dependent contributions include
\begin{equation*}
 m^2\sum_{i=1}^{4}(\partial_i\varphi)^2,\qquad
 m^2\varphi^4,\qquad m^4\varphi^2,
\end{equation*}
Such contributions are treated separately.

As Symanzik also anticipates in his
paper, we introduce an anisotropy with a temporal lattice spacing $a_t$
different from the spatial one $a_s$. 
Distinguishing Euclidean time $\tau=x_4$ from the spatial coordinates 
enlarges the mass-independent basis. 
Temporal reflection and spatial cubic symmetry allow seven structures:
\begin{equation}
 \begin{aligned}
 Q_1&=\sum_{i=1}^{3}(\partial_i^2\varphi)^2,
 &Q_2&=(\nabla^2\varphi)^2,\\
 Q_3&=(\partial_\tau^2\varphi)^2,
 &Q_4&=(\partial_\tau^2\varphi)\nabla^2\varphi,\\
 Q_5&=\varphi^2(\nabla\varphi)^2,
 &Q_6&=\varphi^2(\partial_\tau\varphi)^2,\\
 Q_7&=\varphi^6.
 \end{aligned}
\end{equation}
The relation to the isotropic basis in \cref{eq:isotropic_basis} follows from
\begin{align*}
 \sum_{i=1}^{4}(\partial_i^2\varphi)^2&=Q_1+Q_3,\\
 (\Box_E\varphi)^2&=Q_2+2Q_4+Q_3,\\
 \varphi^2(\partial\varphi)^2_E&=Q_5+Q_6.
\end{align*}
For an isotropic Euclidean lattice, these decompositions imply 
$d_3=d_1+d_2$, $d_4=2d_2$, and $d_6=d_5$, where $d_j$ denotes the 
coefficient of $Q_j$. These relations express the symmetry between temporal 
and spatial directions. They do not generally apply when $a_t\ne a_s$, and 
in particular must not be imposed when comparing with a continuous-time Hamiltonian.
In the anisotropic theory, the coefficients of the 
seven structures are independent before EOM reduction. We write
\begin{equation*}
 \mathcal L_{E,2}=\sum_{j=1}^{7}d_jQ_j
\end{equation*}
for this sector, with explicit mass-dependent operators included separately.
The coefficients $d_j$ depend on the lattice action, the couplings, and the 
anisotropy, and symmetry alone does not determine them. 
We assume that the temporal continuum limit $\xi=a_s/a_t\to\infty$ exists
at fixed $a_s$ and that the result can be expanded in powers of $a_s$. In
the following, $d_j(\xi)$ denotes the coefficient of $Q_j$ at
fixed anisotropy $\xi$, with the explicit powers of $a_s$ factored out as
in \cref{eq:action_spatial_expansion}; it may therefore depend on $\xi$. The reduction to the
Hamiltonian basis, \cref{eq:scalar_ci_di_correspondence} is made at fixed $\xi$. The
continuum-time limit is taken on the resulting $c_i(\xi)$. At
tree level the $d_j$ are polynomials in $1/\xi$ and the limit can be read
off directly. Beyond tree level the limit has to be taken on the matched
$c_i$, since it cannot be inferred from the explicit powers of $a_t$ in
the classical action.

We now derive the classical correspondence between these Euclidean coefficients 
and the coefficients of the reduced Hamiltonian basis. The operators
$Q_1$, $Q_2$, $Q_5$, and $Q_7$ 
contain only spatial derivatives. $Q_3$ and $Q_4$ contain second time derivatives, but 
they can be treated perturbatively at order $a_s^2$.
For $Q_4$, integration by parts in Euclidean time and space gives
\begin{equation}
\begin{aligned}
\int \mathrm{d}\tau\,\mathrm{d}^3x\,Q_4
&=\int \mathrm{d}\tau\,\mathrm{d}^3x\,
  (\partial_\tau^2\varphi)\nabla^2\varphi\\
&=-\int \mathrm{d}\tau\,\mathrm{d}^3x\,
  (\partial_\tau\varphi)\nabla^2(\partial_\tau\varphi)\\
&=\int \mathrm{d}\tau\,\mathrm{d}^3x\,
  \sum_{i=1}^{3}(\partial_i\partial_\tau\varphi)^2,
\end{aligned}
\end{equation}
where boundary terms have been discarded. Thus $Q_4$ can be
represented using only first order time derivatives, while $Q_6$ already
has this property. The remaining operator
$Q_3=(\partial_\tau^2\varphi)^2$ is reduced perturbatively using
the leading Euclidean equation of motion,
$\partial_\tau^2\varphi=(-\nabla^2+m^2)\varphi+\frac{\lambda}{6}\varphi^3$.
The connection between the Euclidean coefficients $d_j$ and the
Hamiltonian coefficients $c_i$ can now be made explicit. At the
classical level, the leading equation of motion reduces $Q_3$ to
$[(-\nabla^2+m^2)\varphi+\lambda\varphi^3/6]^2$. After spatial
integration by parts, its contribution is
\begin{equation}
  \begin{aligned}
    \int \mathrm{d}^3x\,Q_3\simeq\int \mathrm{d}^3x\Big[\,&
      (\nabla^2\varphi)^2+\lambda\varphi^2(\nabla\varphi)^2
      +\frac{\lambda^2}{36}\varphi^6\\
      &+2m^2(\nabla\varphi)^2+m^4\varphi^2
      +\frac{\lambda m^2}{3}\varphi^4\Big].
  \end{aligned}
  \label{eq:scalar_Q3_reduction}
\end{equation}
The symbol $\simeq$ denotes equivalence at this order under 
the leading equation of motion and spatial total derivatives. After continuation 
to Minkowski time and the perturbative Legendre transformation, 
$Q_4$ and $Q_6$ yield $-d_4(\nabla\pi)^2$ and $-d_6\phi^2\pi^2$, respectively. 
Eliminating these momentum dependent operators by the canonical transformations 
described in \cref{sec:Hamiltonian} gives the coefficients of the reduced Hamiltonian basis:
\begin{equation}
\begin{aligned}
c_1&=d_2+d_3-d_4,\\
c_2&=d_1,\\
c_3&=d_5+\lambda d_3-\frac{\lambda}{2}d_4-d_6,\\
c_4&=d_7+\frac{\lambda^2}{36}d_3-\frac{\lambda}{18}d_6.
\end{aligned}
\label{eq:scalar_ci_di_correspondence}
\end{equation}
To compare with \cref{sec:tree}, we expand the tree-level lattice action
in $a_s$ and $a_t$,
$\hat p_i^{\,2}=p_i^2-a_s^2p_i^4/12+\dots$ and
$\hat p_4^{\,2}=p_4^2-a_t^2p_4^4/12+\dots$, so that
\begin{equation}
  \mathcal L_{E,2} = -\frac{1}{24}\,Q_1-\frac{1}{24\,\xi^{2}}\,Q_3 ,
  \label{eq:LE2tree}
\end{equation}
that is, $d_1=-1/24$, $d_3=-1/(24\xi^2)$, and all other $d_j$ vanish
at tree level. Inserting this into \cref{eq:scalar_ci_di_correspondence} at fixed $\xi$ gives
\begin{equation}
  c_1=-\frac{1}{24\xi^{2}},\quad
  c_2=-\frac{1}{24},\quad
  c_3=-\frac{\lambda}{24\xi^{2}},\quad
  c_4=-\frac{\lambda^{2}}{864\,\xi^{2}} ,
  \label{eq:ctree}
\end{equation}
together with the mass-dependent terms
$-\frac{1}{24\xi^{2}}\big[2m^{2}(\nabla\phi)^{2}+m^{4}\phi^{2}
+\tfrac{\lambda}{3}m^{2}\phi^{4}\big]$.
In the continuous-time limit $\xi\to\infty$ only $c_2=-1/24$ survives,
\begin{equation}
  c_1=0,\quad c_2=-\frac{1}{24},\quad c_3=c_4=0 ,
\label{eq:coefficients_anisotropic}
\end{equation}
in agreement with the direct Hamiltonian expansion of
\cref{sec:tree}, including the vanishing of the mass-dependent terms.

For $\xi=1$ we have $d_3=d_1=-1/24$ and $d_2=d_4=d_5=d_6=0$. These
satisfy the isotropy relations $d_3=d_1+d_2$, $d_4=2d_2$, $d_6=d_5$,
so \cref{eq:scalar_ci_di_correspondence} reduces to $c_1=c_2=d_1$
 and $c_3=\lambda d_1$. This is the
form of the on-shell basis for the isotropic lattice. The operator $(\Box_E\phi)^2$ 
is redundant under the leading equation of motion.
 Its elimination shifts the six-field and mass-dependent operators, 
 while its coefficient drops out of $c_1,c_2,c_3$. 
 With the values above one has
\begin{equation}
  c_1=c_2=-\frac{1}{24},\quad c_3=-\frac{\lambda}{24},\quad
  c_4=\lambda^2 \frac{d_3}{36}.
\end{equation}
This describes the isotropic Euclidean lattice. 
In the continuous time limit of the anisotropic case, \cref{eq:coefficients_anisotropic},
there is no lattice symmetry that interchanges temporal and spatial directions, 
so the isotropic relation $c_1=c_2$ is no longer required.

Although terms proportional to positive powers of the temporal lattice 
spacing vanish at tree level as $a_t \rightarrow 0$, radiative corrections 
can generate operator mixing with coefficients depending on the anisotropy 
$\xi$.

\section{Pure Lattice Gauge Theory}
\label{sec:Gauge}

Working again in $d=3$ dimensions on the lattice $\Lambda$, the link
operators 
\begin{equation}
  U_{n,i}\in \mathrm{SU}(N)
\end{equation}
connect sites $n$ and $n+e_i$. We will follow the convention that the
generators $T^a$ ($a=1,\dots,N^2-1$) are Hermitian, with
$\mathrm{tr}\,T^aT^b=\tfrac12\delta^{ab}$ and
$[T^a,T^b]=if^{abc}T^c$. In a non-Abelian lattice gauge theory, there
are left and right Lie derivatives, $L$ and $R$, which act like
\begin{equation}
  \begin{split}
    [L^a_{n,i},U_{m,j}]&=-\delta_{nm}\delta_{ij}\,T^aU_{n,i}\,,\\
    [R^a_{n,i},U_{m,j}]&=+\delta_{nm}\delta_{ij}\,U_{n,i}T^a \,.\\
  \end{split}
\end{equation}
The $L$ and $R$ operators fulfil the algebra
\begin{equation}
  \begin{split}
    [L^a_{n,i},L^b_{m,j}]&=i\delta_{nm}\delta_{ij}f^{abc}L^c_{n,i}\,,\\
    [R^a_{n,i},R^b_{m,j}]&=i\delta_{nm}\delta_{ij}f^{abc}R^c_{n,i}\,,\\
    [L^a_{n,i},R^b_{m,j}]&=0 \,,\\
  \end{split}
\end{equation}
and they are related on each link as
\begin{equation}
  R^bT^b=-\,U_{n,i}^\dagger\,(L^aT^a)\,U_{n,i}
\end{equation}
such that
\begin{equation}
  L^a_{n,i}L^a_{n,i}=R^a_{n,i}R^a_{n,i}\,.
\end{equation}
For the Abelian U(1) theory one has $R=-L$. See
Ref.~\cite{Garofalo:2026pwd} for important consequences of this
difference between Abelian and non-Abelian theories.

With this notation, the KS lattice Hamiltonian reads
\begin{equation}
  \begin{split}
    H_\mathrm{KS}=&\frac{c_0g^2}{2a_s}\sum_{n\in\Lambda}\sum_{i=1}^{3}L^a_{n,i}L^a_{n,i}\;\\
    &+\;\frac{c_0}{g^2a}\sum_{n\in\Lambda}\sum_{1\le i<j\le
      3}\Big[2N-\mathrm{tr}\big(U_{n,ij}+U_{n,ij}^\dagger\big)\Big]\,,\\
  \end{split}
\end{equation}
where
\begin{equation}
  U_{n,ij}=U_{n,i}\,U_{n+e_i,j}\,U^\dagger_{n+e_j,i}\,U^\dagger_{n,j}
\end{equation}
represents the so-called plaquette operator. Summing over $i<j$ counts
each plaquette once. The coefficient $c_0$ is again present due to
breaking of Lorentz invariance~\cite{Carena:2022hpz}, and must in
principle be tuned appropriately for 
instance from the $p^2$ dependence of the lowest glueball mass at
small momenta. In pure gauge theory $c_0$ represents an overall factor
which sets the time unit. Equivalently, one could write two couplings
$g_E$ and $g_M$ with $c_0=g_E/g_M$. 

Because $L^2=R^2$, the electric term can equally
be written in the manifestly symmetric form
$\frac{g^2}{4a}\sum_{n,i}\big(L^a_{n,i}L^a_{n,i}+R^a_{n,i}R^a_{n,i}\big)$. 
The KS Hamiltonian is further constrained by Gauss' law. The Gauss
operators
\begin{equation}
  G^a_n=\sum_{i=1}^3\Big(L^a_{n,i}+R^a_{n-e_i,i}\Big)
\end{equation}
are generators of gauge transformations, satisfying
$[G^a_n,H_\mathrm{KS}]=0$, and physical states obey
$G^a_n|\psi\rangle=0$.

The KS Hamiltonian is formulated in temporal or Weyl gauge, in which
the temporal component of the vector potential $A_0=0$. Therefore, the
continuum counterpart reads
\begin{equation}
  H_\mathrm{YM}\ =\ \int \mathrm{d}^3x\,\tfrac12\big(E^a_iE^a_i+B^a_iB^a_i\big)
\end{equation}
with the first term the chromo-electric, and the second the
chromo-magnetic term.
Here, $\big(A^a_i,\;E^a_i\big)$ is the canonical
pair, with
$[A^a_i(x),E^b_j(y)]=i\delta^{ab}\delta_{ij}\delta(x-y)$.
With $A_0=0$, the relations to the vector potential read
\begin{equation}
  E_i^a = \partial_0 A_i^a\,,\quad B_i^a=\epsilon_{ijk}\partial_j A_k^a+\frac{g}{2}\epsilon_{ijk}f^{abc}A_j^bA_k^c\,,
\end{equation}
or
\begin{equation}
  B_i = \frac{1}{2}\epsilon_{ijk}F_{jk}\,,\quad F_{jk} = \epsilon_{jki}B_i\,.
\end{equation}
We use the convention
\begin{equation}
  A_i = A_i^a T^a\,,\quad E_i = E_i^aT^a\,,\quad B_i = B_i^aT^a\,.
\end{equation}
With $D_i$
the spatial covariant derivative
\begin{equation}
  D_i X=\partial_i X-ig [A_i, X],
\label{eq:covariant_derivative}
\end{equation}
the Gauss law reads
\begin{equation}
  (D_iE_i)^a|\psi\rangle = 0\,.
\end{equation}
To leading order in the lattice spacing $a_s$ the KS Hamiltonian
reduces to the above YM Hamiltonian after the identification
\begin{equation}
  \begin{split}
    L^a_{n,i}&\simeq - \tfrac{a_s^2}{g}E^a_i\,,\\
    U_{n,i}&\simeq e^{ia_sgA_i}\,,\\
    U_{n,ij}&\simeq e^{ia_s^2gF_{ij}}\,.\\
  \end{split}
\end{equation}
Here, $F_{ij}$ are the spatial components of the field strength
tensor. Thus, the building block for the effective Hamiltonian are the
$E_i$ and $B_i$ fields, as well as the covariant derivative $D_i$. The
dimensions are $[E]=[B]=2$ and $[D]=1$.

We may use the following identities to find redundant terms:
$[D_i,D_j]X=-ig\,[F_{ij},X]$, the Bianchi identity $D_{[i}F_{jk]}=0$
(not necessarily for compact U(1)),
and Gauss' law $D_iE_i=0$. Moreover, terms of the form $i[H_0,X]$ are
redundant as discussed previously. Finally, total spatial derivatives
integrate to zero.

The invariant tensors used in the contractions are $\delta_{ij}$ and the 
cubic tensor $\delta_{ijkl}$, together with the invariant pseudo-tensor 
$\epsilon_{ijk}$. The contraction of colour indices is conveniently
performed with traces.

The relevant symmetries are parity $P$, charge conjugation
$C$ and time reversal $T$, which are symmetries also of the lattice
Hamiltonian. The $E$ and $B$ fields, and the derivative $D$ transform
according to the following table

\begin{center}
  \begin{tabular*}{.6\linewidth}{@{\extracolsep{\fill}}llccc}
    \hline
    & $E_i$ & $B_i$ & $D_i$ \\
    \hline\hline
    P & $-$ & $+$ & $-$ \\
    T & $+$ & $-$ & $+$ \\
    C & $-E^T$ & $-B^T$ & covariant\\
    \hline
  \end{tabular*}
\end{center}

At $O(a)$ the operators need to be of dimension five,
which can be obtained by $\tr{DDDX}$, where $X$ can be either $E$ or
$B$. However, since $\tr{T^a}=0$, such combinations vanish. Another
way to obtain dimension five is by combining one derivative with two
fields, and the $\epsilon$ tensor. Possible combinations are
\begin{equation}
  \begin{split}
    \tr{E_i\epsilon_{ijk}D_j E_k}\,,\quad \tr{B_i\epsilon_{ijk}D_j B_k}\,,\\
    \tr{E_i\epsilon_{ijk}D_j B_k}\,,\quad \tr{B_i\epsilon_{ijk}D_j E_k}\,.\\
  \end{split}
\end{equation}
The last two are not independent, their difference is the total
derivative of $\tr{\epsilon_{ijk}E_jB_k}$, such that the integrals over
the two are identical. All three remaining combinations can be made
Hermitian by symmetrising the $E$--$A$ ordering.
The two combinations in the first row are parity odd, and even under charge
conjugation and time reversal. The remaining combination is even under
parity and charge conjugation, but odd under time reversal. 
Therefore, no gauge-invariant dimension-five Hamiltonian density 
compatible with parity and time reversal survives, and $H_1=0$.

Dimension six combinations can be constructed by two $E$ or $B$ fields
accompanied by two derivatives, or by the combination of three $E$ or $B$ fields.
The former must contain either two $E$ or two $B$ fields,
respectively, the mixed combination is excluded by $T$.

The relevant symmetries must now include also $O_h$ and gauge
invariance. The possible operators and their transformations are
listed in \cref{tab:dim6}. After removing all operators forbidden by
symmetry and the redundant ones, one is left with the following list
of Hermitian operators
\begin{equation}
  \begin{split}
    \mathcal{M}_1&\ =\ \sum_{i,j}\mathrm{tr}(D_iF_{ij}D_iF_{ij})\,,\\
    \mathcal{M}_3&\ =\ \mathrm{tr}(D_iF_{ik}D_jF_{jk})\,,\\
    \mathcal{E}_1&\ =\ \sum_i\mathrm{tr}(D_iE_iD_iE_i)\,,\\
    \mathcal{K}_1&\ =\ i\,\mathrm{tr}(F_{ij}F_{jk}F_{ki})\,,\\
    \mathcal{K}_2&\ =\ i\,\mathrm{tr}(E_iE_jF_{ij})\,.\\
  \end{split}
\label{eq:list_hermitian_operators}
\end{equation}
The basis is understood modulo spatial total derivatives in the 
integrated Hamiltonian, terms proportional to the Gauss constraint, 
and gauge-invariant local canonical transformations. In particular, 
$\mathcal E_1$ contains no sum inside each square and is distinct 
from the square of the summed Gauss constraint. It is therefore 
not eliminated by Gauss’ law.

The dimension-six contribution to the effective Hamiltonian can therefore
be written as
\begin{equation}
H_2=\int \mathrm{d}^3x\,
\left[
c_{\mathrm{}1}\mathcal M_1+
c_{\mathrm{}2}\mathcal M_3+
c_{\mathrm{}3}\mathcal E_1+
c_{\mathrm{}4}\mathcal K_1+
c_{\mathrm{}5}\mathcal K_2
\right].
\label{eq:gauge-general-H2}
\end{equation}
The coefficients are determined by matching to the specified lattice
Hamiltonian. Products involving canonical electric fields are understood
with a Hermitian ordering prescription. In particular, $\mathcal K_2$ 
is the Hermitian part of $i\operatorname{tr}(E_i E_j F_{ij})$.
For comparison, Lüscher and Weisz classify dimension-six operators in
the isotropic Euclidean theory~\cite{Luscher:1984xn}. In the following Euclidean comparison, 
$i,j,k$ run from 1 to 4, with $x_4=\tau$. Elsewhere in this section, they denote 
spatial components and run from 1 to 3. We use anti-Hermitian
fields,
\begin{equation}
\mathcal A_i=-igA_i^{\mathrm E},\qquad
\mathcal F_{ij}=-igF_{ij}^{\mathrm E},\label{eq:gauge-Euclidean-conventions}
\end{equation}
Their operator basis is built from these anti-Hermitian fields:
\begin{align}
O_1&=\sum_{i,j}\operatorname{tr}
\left[(\mathcal D_i\mathcal F_{ij})
      (\mathcal D_i\mathcal F_{ij})\right],\nonumber\\
O_2&=\sum_{i,j,k}\operatorname{tr}
\left[(\mathcal D_i\mathcal F_{jk})
      (\mathcal D_i\mathcal F_{jk})\right],\nonumber\\
O_3&=\sum_{i,j,k}\operatorname{tr}
\left[(\mathcal D_i\mathcal F_{ik})
      (\mathcal D_j\mathcal F_{jk})\right].
\label{eq:gauge-Euclidean-basis}
\end{align}
Only the displayed sums are taken in these equations. The trace is taken
in the fundamental representation, with the same normalisation as in
the Hamiltonian basis. The factor $-g^2$ arises from the two field strengths, 
and we define
\begin{equation}
\widetilde O_r=-\frac{1}{g^2}O_r,\qquad r=1,2,3.
\end{equation}

Table II displays the decomposition of the Euclidean operators in \cref{eq:gauge-Euclidean-basis} 
into temporal and spatial components, together with their analytically continued forms.
We use $t=-i \tau$ and $F_{4i}^{E}\rightarrow -iE_i$, and the
leading Hamilton equations needed for the reduction are
\begin{equation}
\partial_0 E_i=D_jF_{ji},\qquad
\partial_0 F_{ij}
=D_i E_j-D_j E_i.
\label{eq:gauge-leading-evolution}
\end{equation}
For the reduction, we use the operators as given in Table I
\begin{align}
\mathcal M_2&=\operatorname{tr}
(D_i F_{j k}D_i F_{j k}),\nonumber\\
\mathcal E_2&=\operatorname{tr}
(D_i E_j D_i E_j),\nonumber\\
\mathcal E_4&=\operatorname{tr}\
\left(D\cdot E\right)^2
\label{eq:gauge-additional-densities}
\end{align}
And then
\begin{align}
\widetilde O_1
&\rightarrow M_1-\mathcal E_1+\mathcal M_3,\nonumber\\
\widetilde O_2
&\rightarrow
\mathcal M_2-4\mathcal E_2+2\mathcal M_3 +4g\mathcal K_2+2\mathcal E_4,\nonumber\\
\widetilde O_3
&\rightarrow 0.
\label{eq:gauge-continued-basis}
\end{align}
The last line follows from an equation-of-motion reduction.

The Bianchi identity and integration by parts imply

\begin{equation}
\int \mathrm{d}^3x\,\mathcal M_2
=\int \mathrm{d}^3x\,(2\mathcal M_3+4g\mathcal K_1).
\label{eq:gauge-magnetic-reduction}
\end{equation}
The electric reduction follows from the gauge-invariant classical
generator
\begin{equation}
\Phi_3=\int \mathrm{d}^3x\,
\operatorname{tr}(E_i D_j F_{ji}),
\end{equation}
whose leading time evolution is
\begin{equation}
\frac{\mathrm{d}\Phi_3}{\mathrm{d}t}=\int \mathrm{d}^3x\,
(\mathcal M_3-\mathcal E_2+\mathcal E_4+4g\mathcal K_2).
\label{eq:gauge-electric-reduction}
\end{equation}
The elimination of $\mathcal E_2$ can be made explicit through a local
canonical transformation.
In the operator formulation, the corresponding
time evolution is $\dot \Phi_3=i[H_0,\Phi_3]$, with a consistent Hermitian ordering
of the generator and composite operators. Consequently, a contribution
$a_s^2\int \mathrm{d}^3x\,\mathcal E_2$ can be expressed as
\begin{equation}
a_s^2\int \mathrm{d}^3x\,
\left(\mathcal M_3+\mathcal E_4+4g\mathcal K_2\right)
-a_s^2\,i[H_0,\Phi_3].
\end{equation}
The transformation $H'=UHU^\dagger$ with
$U=\exp(-ia_s^2\Phi_3)$ adds $a_s^2\,i[H_0,\Phi_3]$ through order $a_s^2$,
thereby cancelling the commutator term.
Using Gauss law to remove $\mathcal E_4$ results in
\begin{equation}
\int \mathrm{d}^3x\,\mathcal E_2\simeq
\int \mathrm{d}^3x\,\left(\mathcal M_3+4g\mathcal K_2\right).
\label{eq:gauge-E2-equivalence-short}
\end{equation}
Here $\simeq$ denotes equivalence under a local canonical transformation
and the Gauss constraint, rather than equality in the original variables.
The transformation shifts the coefficients of $\mathcal M_3$ and
$\mathcal K_2$ and preserves the spectrum through the retained order;
other observables must be transformed consistently.

Substituting \cref{eq:gauge-E2-equivalence-short} into \cref{eq:gauge-continued-basis}
gives the combination $\mathcal M_2-2\mathcal M_3-12g\mathcal K_2$
 on the physical subspace. Using \cref{eq:gauge-magnetic-reduction}, the two contributions proportional to $\mathcal M_3$ cancel.

Thus, modulo the Gauss constraint and a local canonical transformation,
\begin{equation}
\int \mathrm{d}^3x\;\widetilde O_2
\rightarrow \int \mathrm{d}^3x\, 4g (\mathcal K_1-3\mathcal K_2).
\label{eq:gauge-cubic-reduction}
\end{equation}
In particular, $\mathcal K_1$ enters through the reduction of
$\mathcal M_2$.

These identities describe the classical reduction of analytically
continued operator densities. Matching their coefficients to a
Hamiltonian additionally requires the Legendre transformation and a
specified choice of canonical variables.

\subsection{Tree-Level Matching}
We determine the quadratic part of the tree-level matching of the Kogut-Susskind Hamiltonian 
by expanding about the trivial gauge-field configuration. At this order the colour components 
decouple, and the calculation reduces to the Abelian theory.
Let us compare this to the tree-level result, first derived in
Ref.~\cite{Luo:1998dx}. For completeness, we re-derive the result
following Ref.~\cite{Luscher:1984xn} and take the link as the parallel
transporter along the link
\begin{equation}
  U_{n,i} = \mathcal P \exp\left(ig\int_{a_sn}^{a_sn + a_se_i}A_i \mathrm{d}x_i \right)\,.
\end{equation}
Let
\begin{equation}
  U_{n,i} = e^{i\theta_{n,i}} = e^{ia_sgA_{n,i}^\mathrm{lat}}\,,
\end{equation}
such that
\begin{equation}
  \theta_{n,i} = a_sgA_{n,i}^\mathrm{lat}\,,
\end{equation}
and
\begin{equation}
  [A_{n,i}^\mathrm{lat}, E_{m,j}^\mathrm{lat}] = \frac{i}{a^3}\delta_{m,n}\delta_{i,j}\,.
\end{equation}
For the plaquette angle $\theta_{ij}^p$ we have to first non-trivial order
\begin{equation}
  \label{eq:loopContr}
  1 - \cos\theta_{ij}^p = \frac{1}{2}|\theta_{ij}^p|^2\,.
\end{equation}
Define
\begin{equation}
  \sigma_i = \frac{\hat{p}_i}{p_i} = 1 - \frac{a_s^2p_i^2}{24} + O(a_s^4)\,,
\end{equation}
with as before
\begin{equation}
  \hat{p}_i = \frac{2}{a_s}\sin(p_i a_s/2) = p_i\sigma_i\,.
\end{equation}
The free $A$ field in momentum space is the line average of the
continuum field along the link due to the identification of the link
with the parallel transporter. Therefore,
\begin{equation}
  A_i^\mathrm{lat}(\mathbf{p}) = \sigma_iA_i(\mathbf{p})
\end{equation}
(no sum over $i$ on the right hand side), and to preserve the
commutation relations
\begin{equation}
  E_i^\mathrm{lat}(\mathbf{p}) = \sigma_i^{-1} E_i(\mathbf{p})\,.
\end{equation}
In the Abelian theory, these $E$ fields are related to the generators
\begin{equation}
  L = -\frac{a_s^2}{g}E^\mathrm{lat}\,,\quad R = -L\,.
\end{equation}
If we also define
\begin{equation}
  F_{ij}^\mathrm{lat} = \Delta_iA_{n,j}^\mathrm{lat} - \Delta_j A_{n,i}^\mathrm{lat}\,,
\end{equation}
where $\Delta_i$ is the forward lattice derivative in the spatial direction $i$.
We write the KS lattice Hamiltonian as
\begin{equation}
  H_\mathrm{KS} = a_s^3\sum_n\left[\frac{1}{2}\sum_i\left( E_{n,i}^\mathrm{lat}\right)^2 + \frac{1}{2}\sum_{i<j}(F_{n,ij}^\mathrm{lat})^2\right]\,,
\end{equation}
which can be Fourier transformed in the standard way to read in
continuum variables
\begin{equation}
  \tilde{H}_{\rm KS}=\int\!\frac{\mathrm{d}^3p}{(2\pi)^3}\Big[\tfrac12\sum_i\sigma_i^{-2}|E_i|^2+\tfrac12\sum_{i<j}\sigma_i^2\sigma_j^2|F_{ij}|^2\Big]\,.
\end{equation}
Now, expand in $a_s$
\begin{equation}
  \sigma_i^2\sigma_j^2\,|F_{ij}|^2 = |F_{ij}|^2\left[1 - \frac{a_s^2}{12}(p_i^2 + p_j^2)\right] + O(a^4)
\end{equation}
such that the magnetic energy is
\begin{equation}
  \frac{1}{2}\left[\sum_{i<j}|F_{ij}|^2 -
  \frac{a_s^2}{12}\sum_{i,j}|p_i|^2|F_{ij}|^2\right] + O(a_s^4)\,.
\end{equation}
For the electric energy one finds
\begin{equation}
  \begin{split}
    \frac{1}{2}|E_i^\mathrm{lat}|^2 =&
    \frac{1}{2}\sigma_i^{-2}|E_i|^2\\
    =& \frac{1}{2} |E_i|^2\left[1 + \frac{a_s^2p_i^2}{12}\right] + O(a_s^4)\,.
  \end{split}
\end{equation}
Transforming back to position space and retaining the quadratic parts 
of the operators in \cref{eq:list_hermitian_operators}, we use
\begin{equation}
\mathcal E_1=\frac12\sum_{a,i}
(\partial_i E_i^a)^2,\qquad
\mathcal M_1=\frac12\sum_{a,i,j}
(\partial_i F_{ij}^a)^2,
\end{equation}
where only the explicit sums are taken, we obtain
\begin{equation}
H_{\mathrm{eff}}^{}=H_{YM}
+a_s^2\int \mathrm{d}^3x\left[-\frac1{12}\mathcal M_1^{}
+\frac1{12}\mathcal E_1^{}\right]+O(a_s^4).
\label{eq:gauge-quadratic-matching-result}
\end{equation}
This matching does not determine the independent coefficients
of $\mathcal K_1$ and $\mathcal K_2$, which have no quadratic parts.
Their absence from
\cref{eq:gauge-quadratic-matching-result} does not establish
their absence from the full interacting effective Hamiltonian.

The present calculation establishes the quadratic tree-level matching.
Determining the independent interaction coefficients in
\cref{eq:gauge-general-H2} requires non-Abelian matching in the
same canonical representation, using a covariant expansion of the
lattice Hamiltonian or suitable interaction amplitudes. This provides
a direct extension of the quadratic analysis.

\section{Conclusion}

In this paper we have proposed a framework to construct the Symanzik
effective Hamiltonian for Hamiltonian lattice field
theories. 
Analogously to the Symanzik effective action, contributions
to the effective Hamiltonian $H_\mathrm{eff}$ at $O(a_s^k), k\geq0$ 
in the spatial lattice spacing must have dimension $d+1+k$,
respect the symmetries of the theory, and contain only spatial
derivatives. 
One important point is that 
Lorentz invariance is broken down to $O_h$, which needs to be
incorporated in the classification process. 
The effective Hamiltonian acts on the continuum Hilbert
space. Independent on-shell bases are obtained by eliminating
redundancies,  
including spatial total derivatives and terms generated by commutators 
with the leading continuum Hamiltonian. The latter correspond to perturbative 
canonical transformations, with observables transformed consistently.

We have also examined the connection to the corresponding anisotropic 
Symanzik effective action through the temporal continuum limit and a 
perturbative Legendre transformation.  This provides a complementary route to 
the Hamiltonian description once higher order time derivatives have been treated.
Note that the Hamiltonian limit has been tested non-perturbatively in
Refs.~\cite{Funcke:2022opx,Gross:2025qae} in $2+1$ dimensional U(1)
lattice gauge theory.

We have applied the framework to $\phi^4$ and pure lattice gauge theory, 
constructing operator bases to order $a_s^2$. Under the assumptions of this work, 
the classification excludes corrections linear in $a_s$ to all orders in perturbation theory. 
For the scalar theory, we have established the correspondence with the effective action 
and verified the tree-level matching. For pure gauge theory, we have related the effective action 
operators to the Hamiltonian basis and reproduced the known quadratic tree-level corrections 
to the Kogut–Susskind Hamiltonian.

Earlier studies constructed improved lattice gauge Hamiltonians from
improved actions~\cite{Luo:1998dx} and directly within the Hamiltonian 
formulation~\cite{Carlsson:2001wp}. The present work complements 
these constructions by organising lattice artefacts in a continuum effective 
Hamiltonian, with explicit operator classification and reduction to 
independent on-shell bases. Its connection to the effective action also 
relates to the treatment of higher order time derivatives in effective Lagrangians 
discussed by Grosse-Knetter~\cite{Grosse-Knetter:1993tae}.

The framework developed here forms a basis for an improvement 
programme starting from a given lattice Hamiltonian. 
The direct Hamiltonian construction avoids dealing with higher order time derivatives, 
while the effective action route required their prior treatment. The development of
non-perturbative improvement conditions, the completion of the non-Abelian interaction 
matching, and applications to theories with fermions are subject
for future research.

\begin{table*}[t]
  \centering
  \caption{Gauge-invariant dimension-6 candidate operators for $H_2$ in pure
    $SU(N)$ gauge theory, built from $E_i$, $B_i$ ($F_{ij}=\epsilon_{ijk}B_k$)
    and $D_i$. No summation is implied inside the squares of the cubic
    operators. Transformation rules: $P$: $E_i\to-E_i$, $B_i\to B_i$,
    $D_i\to-D_i$; $T$ (antiunitary): $E_i\to E_i$, $B_i\to-B_i$, $i\to-i$;
    $C$: $E_i\to-E_i^T$, $B_i\to-B_i^T$, reversing the order in traces.}
  \label{tab:dim6}
  \begin{tabular*}{\textwidth}{@{\extracolsep{\fill}}llcccccl}
\hline
    operator & structure & $O_h$ & $P$ & $T$ & $C$ & $CPT$ & status \\
    \hline
    $\mathcal{M}_1=\sum_{i,j}\mathrm{tr}(D_iF_{ij}D_iF_{ij})$ & magnetic & cubic & $+$ & $+$ & $+$ & $+$
      & allowed; tree-level KS: $-\tfrac1{12}$ \\
    $\mathcal{M}_2=\mathrm{tr}(D_iF_{jk}D_iF_{jk})$ & magnetic & rot. & $+$ & $+$ & $+$ & $+$
      & Bianchi: $\mathcal{M}_2=2\mathcal{M}_3+ 4g \mathcal{K}_1$ \\
    $\mathcal{M}_3=\mathrm{tr}(D_iF_{ik}D_jF_{jk})$ & magnetic & rot. & $+$ & $+$ & $+$ & $+$
      & allowed \\
    $\mathcal{M}_4 =\mathrm{tr}(D_iB_jD_jB_i)$ & magnetic & rot. & $+$ & $+$ & $+$ & $+$
      & reduces via $D\cdot B=0$ to $2g\mathcal{K}_1$ \\
    $\mathcal{E}_1=\sum_i\mathrm{tr}(D_iE_iD_iE_i)$ & electric & cubic & $+$ & $+$ & $+$ & $+$
      & allowed; tree-level KS: $+\tfrac1{12}$ \\
    $\mathcal{E}_2=\mathrm{tr}(D_iE_jD_iE_j)$ & electric & rot. & $+$ & $+$ & $+$ & $+$
      & redundant: $i[H_0,X]\simeq \mathcal{M}_3-\mathcal{E}_2+4g\mathcal{K}_2$ \\
    $\mathcal{E}_3=\mathrm{tr}(D_iE_jD_jE_i)$ & electric & rot. & $+$ & $+$ & $+$ & $+$
      & reduces to $(D\cdot E)^2+2g\mathcal{K}_2$ \\
    $\mathcal{E}_4=\mathrm{tr}(D\cdot E)^2$ & electric & rot. & $+$ & $+$ & $+$ & $+$
      & vanishes on physical states (Gauss) \\
    $\mathcal{K}_1=i\,\mathrm{tr}(F_{ij}F_{jk}F_{ki})$ & $BBB$ & rot. & $+$ & $+$ & $+$ & $+$
      & allowed \\
    $\mathcal{K}_2=i\,\mathrm{tr}(E_iE_jF_{ij})$ & $EEB$ & rot. & $+$ & $+$ & $+$ & $+$
      & allowed \\
    \midrule
    $\mathrm{tr}(D_iE_jD_iB_j)$, $\mathrm{tr}(D_iE_jD_jB_i)$ & mixed & rot. & $-$ & $-$ & $+$ & $+$
      & forbidden by $P$, $T$ \\
    $\sum_i\mathrm{tr}(D_iE_iD_iB_i)$ & mixed & cubic & $-$ & $-$ & $+$ & $+$
      & forbidden by $P$, $T$ \\
    $i\,\epsilon_{ijk}\mathrm{tr}(E_iE_jE_k)$ & $EEE$ & rot. & $-$ & $-$ & $+$ & $+$
      & forbidden by $P$, $T$ \\
    $i\,\epsilon_{ijk}\mathrm{tr}(E_iB_jB_k)$ & $EBB$ & rot. & $-$ & $-$ & $+$ & $+$
      & forbidden by $P$, $T$ \\
\hline
  \end{tabular*}
\end{table*}

\begin{table*}[t]
  \caption{Temporal and spatial decomposition of the Euclidean gauge
    operators. Definitions use the anti-Hermitian fields
    $\mathcal F_{ij}=-igF_{ij}^{\mathrm E}$ and covariant derivatives
    $\mathcal D_i$ of \cref{eq:covariant_derivative}. Spatial indices $i,j,k$ run from 1 to 3;
    $4$ denotes Euclidean time. All rows are local densities, including
    $T_E$. The third column counts explicit $\mathcal D_4$ factors; it
    does not count derivatives implicit in the field strengths. The last
    column gives the continued forms or their classical reductions,
    understood after spatial integration with periodic boundary conditions.
    The leading Hamilton equations are used where indicated, and the
    Gauss reduction applies on the physical subspace.}
  \label{tab:gauge_temporal_spatial_split}
  \centering
  \begin{tabular*}{.8\textwidth}{@{\extracolsep{\fill}}llcl}
    \hline
    & Euclidean density
    & Explicit $\mathcal D_4$ factors
    & Continued form / reduction \\[5pt]
    \hline
    $S_1$ &
    $\displaystyle\sum_{i,j=1}^{3}\operatorname{tr}
        [(\mathcal D_i\mathcal F_{ij})(\mathcal D_i\mathcal F_{ij})]$
        & $0$ & $-g^2\mathcal M_1$ \\[7pt]
        
        $S_2$ &
        $\displaystyle\sum_{i=1}^{3}\operatorname{tr}
        [(\mathcal D_i\mathcal F_{i4})(\mathcal D_i\mathcal F_{i4})]$
        & $0$ & $g^2\mathcal E_1$ \\[7pt]
        
        $S_3$ &
        $\displaystyle\sum_{j=1}^{3}\operatorname{tr}
        [(\mathcal D_4\mathcal F_{4j})(\mathcal D_4\mathcal F_{4j})]$
        & $2$ & $-g^2\mathcal M_3$ (EOM) \\[7pt]
        
        $S_4$ &
        $\displaystyle\sum_{i,j,k=1}^{3}\operatorname{tr}
        [(\mathcal D_i\mathcal F_{jk})(\mathcal D_i\mathcal F_{jk})]$
        & $0$ & $-g^2\mathcal M_2$ \\[7pt]
        
        $S_5$ &
        $\displaystyle\sum_{i,j=1}^{3}\operatorname{tr}
        [(\mathcal D_i\mathcal F_{4j})(\mathcal D_i\mathcal F_{4j})]$
        & $0$ & $g^2\mathcal E_2$ \\[7pt]
        
        $S_6$ &
        $\displaystyle\sum_{j,k=1}^{3}\operatorname{tr}
        [(\mathcal D_4\mathcal F_{jk})(\mathcal D_4\mathcal F_{jk})]$
        & $2$ &
        $2g^2(\mathcal E_2-\mathcal E_3)$\newline
        $=2g^2(\mathcal E_2-\mathcal E_4-2g\mathcal K_2)$ \\[7pt]
        
        $S_7$ &
        $\displaystyle\sum_{i,j=1}^{3}\operatorname{tr}
        [(\mathcal D_i\mathcal F_{i4})(\mathcal D_j\mathcal F_{j4})]$
        & $0$ & $g^2\mathcal E_4\simeq0$ (Gauss) \\[7pt]
        
        $S_8$ &
        $\displaystyle\sum_{i,j,k=1}^{3}\operatorname{tr}
        [(\mathcal D_i\mathcal F_{ik})(\mathcal D_j\mathcal F_{jk})]$
        & $0$ & $-g^2\mathcal M_3$ \\[7pt]
        
        $S_9$ &
        $\displaystyle\sum_{j,k=1}^{3}\operatorname{tr}
        [(\mathcal D_4\mathcal F_{4k})(\mathcal D_j\mathcal F_{jk})]$
        & $1$ & $g^2\mathcal M_3$ (EOM) \\[7pt]
        
        $T_E$ &
        $\displaystyle\sum_{j,k=1}^{3}\operatorname{tr}
        (\mathcal F_{4k}[\mathcal F_{jk},\mathcal F_{4j}])$
        & $0$ & $-2g^3\mathcal K_2$ \\[7pt]
        \hline
  \end{tabular*}
  \par\smallskip
  \begin{minipage}{\textwidth}
    \small
    Only the displayed sums are taken in the definitions. The trace has
    the same fundamental-representation normalization as in the text.
    The Euclidean operators decompose as
    \[
    \begin{aligned}
      O_1&=S_1+S_2+S_3,\\
      O_2&=S_4+2S_5+S_6+2S_3,\\
      O_3&=S_7+S_8+S_3+2S_9.
    \end{aligned}
    \]
    Here $\mathcal E_3=\sum_{i,j=1}^{3}\operatorname{tr}
    [(D_iE_j)(D_jE_i)]$, and
    $\int d^3x\,\mathcal E_3
    =\int d^3x\,(\mathcal E_4+2g\mathcal K_2)$ at the classical level.
    The factor $-1/g^2$ defining $\widetilde O_r$ is not included in the
    last column. Applying it to these decompositions gives the reductions
    displayed in the text.
  \end{minipage}
\end{table*}
 
\begin{acknowledgments}
  We thank Marco Garofalo, Thomas Luu and Ulf-G. Meißner for helpful
  comments to the manuscript.
  
  This project was funded by the Deutsche Forschungsgemeinschaft (DFG,
  German Research Foundation) as a project in the CRC 1639 NuMeriQS --
  project no.\ 511713970, and under Germany’s Excellence Strategy –
  EXC 3107 – Project-ID 533766364 in the Color-meets-flavor cluster of
  excellence.

  We have used Claude AI for checking the consistency of the
  manuscript and for performing some of the basic calculations.
\end{acknowledgments}

\end{document}